\documentclass[submitted]{eptcs} 
\providecommand{\event}{QPL 2011}
\usepackage{amsfonts, amssymb, amsmath}

\renewcommand{\bar}{\overline}

\newcommand{\1}{{\bf 1}}
\newcommand{\C}{{\cal C}}  
\newcommand{\A}{{\frak A}} 
\newcommand{\B}{{\frak B}} 
\newcommand{\E}{{\bf E}}   
\newcommand{\EC}{{\bf E}({\cal C})} 
\renewcommand{\H}{{\bf H}}   

\newcommand{\Hom}{\text{Hom}}

\usepackage{amsthm}
\theoremstyle{definition}
\newtheorem*{definition}{Definition}
\theoremstyle{theorem}
\newtheorem{theorem}{Theorem}

\title{Symmetry and Self-Duality in Categories of Probabilistic Models}
\author{Alexander Wilce}
\def\titlerunning{Symmetry and Self-Duality in Categories of Probabilistic Models}
\def\authorrunning{Alexander Wilce}
\begin{document}
\maketitle

\begin{abstract}
  This note adds to the recent spate of derivations of the
  probabilistic apparatus of finite-dimensional quantum theory from
  various axiomatic packages. We offer two different axiomatic
  packages that lead easily to the Jordan algebraic structure of
  finite-dimensional quantum theory. The derivation relies on the
  Koecher-Vinberg Theorem, which sets up an equivalence between
  order-unit spaces having homogeneous, self-dual cones, and formally
  real Jordan algebras.
\end{abstract}

\section{Introduction and Overview} 

The last several years have seen a spate of derivations of the probabilistic apparatus of finite-dimensional quantum theory from various axiomatic packages, many having an information-theoretic motivation \cite{CDP,  Dakic-Brukner, Hardy, MM, Rau}. This note (which in part echoes, but greatly improves upon \cite{W11}) adds to the flow.  I offer two different, though overlapping, axiomatic packages, both stressing symmetry principles, that lead quickly and easily to the Jordan algebraic structure of finite-dimensional quantum theory. Quickly and easily, 
at any rate, if one is familiar with the Koecher-Vinberg Theorem  \cite{K, V}, which sets up an equivalence between order-unit spaces having homogeneous, self-dual cones, and formally real Jordan algebras. 

A probabilistic system can be described, in a standard way, in terms of an order-unit space $A$, the positive elements of which are scalar multiples of ``effects". The strategy, then, is to show that certain strong, but not unreasonable, assumptions force the positive cone $A_+$ to be homogeneous and self-dual, and hence, isomorphic to the cone of squares of such a Jordan algebra. In \cite{BDW}, several conditions are adduced that lead to a homogeneous and {\em weakly} self-dual cone --- that is, a homogeneous cone that is {\em order-isomorphic} to its dual cone in $A^{\ast}$.  However, proper self-duality is a much more stringent condition, requiring that the isomorphism be mediated by an inner product. 

The line of attack here is to assume that systems individually have a great deal of symmetry, and collectively, can be organized into a symmetric monoidal category \cite{AC, Ba, Se}.  Here is a sketch. Further details can be found in the longer paper \cite{longer}.

\section{Probabilistic Models}  

A {\em test space} \cite{W10test} is a pair $(X,\A)$ where $X$ is a set of {\em outcomes} and $\A$ is a covering 
of $X$ by non-empty (for our purposes here, finite) subsets called {\em tests}, each understood as the set of possible outcomes of some measurement, experiment, etc. Two outcomes $x, y \in X$ are {\em distinguishable} iff they belong to a common test. In this case, I write $x \perp y$. Notice that there is, as yet, no linear structure in view, let alone an inner product; so this notation is promissory.

 A {\em state} on a test space $(X,\A)$ is a function $\alpha : X \rightarrow [0,1]$,
summing independently to unity on each test. A {\em symmetry} of $(X,\A)$ is a mapping $g : X \rightarrow X$ 
such that $g(E), g^{-1}(E) \in \A$ for every $E \in \A$. By a {\em probabilistic model}, I mean a structure $(X,\A,\Omega,G)$ where $(X,\A)$ is a test space, $\Omega$ is a compact convex set of states on $(X,\A)$, and $G$ is a group acting on $(X,\A)$ by symmetries, and leaving $\Omega$ invariant. 

For illustration, if $\H$ is a finite-dimensional Hilbert space (real or Complex), the 
corresponding {\em quantum model} is $(X(\H), \A(\H), \Omega(\H), U(\H))$, where 
 $X = X(\H)$ is the set of rank-one projection operators on $\H$,  ${\cal A} = \A(\H)$ is set of (projective) {\em frames}, i.e., maximal pairwise orthogonal sets of projections, $\Omega(\H)$ is the convex set of density operators 
on $\H$, and $U(\H)$ is the group of unitary operators on $\H$, acting on $X(\H)$ by conjugation. 

\paragraph{Categories of Models.} I will be interested in categories of models.  A {\em morphism} from a model $(X,\A,\Omega,G)$ to a model $(Y,\B,\Gamma,H)$ is a pair $(\phi,\psi)$, where  
\begin{enumerate} 
\item[(i)] $\phi : X \rightarrow Y$ with $\phi(\A) \subseteq \B$, $\phi^{\ast}(\Gamma) \subseteq \Omega$
\item[(ii)] $\psi \in \Hom(G,H)$; 
\item[(iii)] $\phi(gx) = \psi(g)\phi(x)$ for all $x \in X, g \in G$. 
\end{enumerate} 
In what follows, $\C$ is a symmetric monoidal category of probabilistic models $A = (X(A),\A(A),\Omega(A), G(A))$, with morphisms as above. I shall make two further assumptions: 

\begin{enumerate}
\item[(1)] For every $A \in \C$, $G(A) \subseteq \C(A,A)$. 
\item[(2)] The model $A \otimes B \in \C$ is a {\em composite} of the models $A, B \in \C$, in the sense of \cite{BW}. This means, in particular, that there are canonical injections $ \otimes : X(A) \times X(B) \rightarrow X(A \otimes B)$ and $\otimes : \Omega(A) \times \Omega(B) \rightarrow \Omega(A \otimes B)$, with 
\[E \otimes F = \{ x \otimes y | x \in E, y \in F\} \in \A(A \otimes B)\] for 
every $E \in \A(A), F \in \A(B)$, and 
\[(\alpha \otimes \beta)(x \otimes y) = \alpha(x) \beta(y)\] for all $\alpha \in \Omega(A), \beta \in \Omega(B)$, $x \in X(A)$ and $y \in X(B)$. A {\em bipartite state} between $A, B \in \C$ is a state $\omega$ in $\Omega(A \otimes B)$. It is also part of the definition that the {\em un-normalized conditional state} 
$\hat{\omega}(x) := \omega(x, \cdot)$ belong to $\Omega(B)$ for every $x \in X$, and similarly with $A$ and 
$B$ reversed. 
\end{enumerate}

\paragraph{Models Linearized.} Every model $A  = (X(A),\A(A),\Omega(A),G(A)) \in \C$ generates, in a standard and quite canonical way, an order-unit space $\E(A)$. To be precise, $\E(A)$ is the span in ${\Bbb R}^{\Omega}$ of the evaluation functionals associated with measurement outcomes $x \in X$.)  In the case of a quantum model $A(\H) = (X(\H), \A(\H), \Omega(\H), U(\H))$, one has $\E(A) \simeq {\cal L}(\H)$, the space of Hermitian operators on $\H$, with the usual ordering and $u_A = \1_{\H}$. 

The construction $A \mapsto \E(A)$ is functorial, so we obtain from $\C$ a category $\EC$ of order-unit spaces and positive linear mappings. It is natural to enlarge this to a category ${\cal E}$ in 
which each hom-set ${\cal E}(A,B)$ is an ordered linear space, and in which, e.g., ${\cal E}(I,A) \simeq \E(A)$. In what follows, I assume that such a ``linearized" category $\cal E$ has been 
fixed. 

\section{Bi-Symmetric Models} 

To tighten this structure further, I now ask that every $A \in \C$ enjoy a property I call {\em bi-symmetry}. 

\begin{definition} A model $A \in \C$ is {\em bi-symmetric} iff 
\begin{itemize} 
\item[(i)] $G(A)$ acts transitively on the pure states (that is, extreme points) of $\Omega(A)$, 
\item[(ii)] $G(A)$ acts transitively on $\A$, and on pairs $(x,y)$ of outcomes with $x \perp y$.
\end{itemize} 
If, in place of (ii), we require that arbitrary bijections $f : E \rightarrow F$, $E, F \in \A$, extend to elements of $G$, then $A$ is {\em fully bi-symmetric}. 
\end{definition}

If $A$ is bi-symmetric, then $G$ acts transitively.
Clearly, the quantum model discussed above is fully bi-symmetric. Bi-symmetry and full bi-symmetry, 
are very natural conditions. (See \cite{W09} for further discussion and motivation of the latter.) 

\begin{definition}
 A {\em SPIN form}\footnote{This is probably not the best choice of terminology.} for 
the model $A$ is a real bilinear form $B$ on $\E(A)$ that is {\em symmetric}, {\em positive} in the sense that 
$B(a,b) \geq 0$ for all $a, b \in \E(A)_+$, {\em invariant}, in the sense that $B(ga,gb) = B(a,b)$ for all 
$g \in G(A)$, and {\em normalized}, in the sense that $B(u_A,u_A) = 1$. A SPIN form is {\em orthogonalizing} iff 
$B(x,y) = 0$ for all distinguishable measurement outcomes $x, y \in X(A)$. 
\end{definition}

An example is the usual tracial inner product on ${\cal L}(\H)$. Call $\E(A)$ {\em irreducible} iff (with respect to any SPIN form $B$), the subspace $u^{\perp} = \{ a \in \E(A) | B(a,u) = 0\}$ (this is independent of $B$) is irreducible under the group $G(A)$. Quantum models are irreducible in this sense. 

\begin{theorem}\label{thm-1} If $\E(A)$ is irreducible, it supports at most one orthogonalizing SPIN form, which --- if 
it exists --- is an inner product.
\end{theorem}

\section{Conjugates and Daggers} 

At this point, the aim is to find sufficient conditions for the existence of an orthogonalizing SPIN form on $\E(A)$. I shall provide two. 

\begin{definition} By a {\em conjugate} for a model $A$, I mean a structure $(\overline{A},\gamma_{A}, \eta_{A})$, where $\gamma_{A} :  A \mapsto \overline{A}$ is an isomorphism of models, and $\eta_{A}$ is a 
bipartite state on $A \times \overline{A}$ such that $\eta(x,\gamma_{A}(x)) = 1/n$ (where $n$ is the rank of $A$)  for every $x \in X(A)$. 
\end{definition}

In the case of a quantum-mechanical model $A = A(\H)$ associated with a Hilbert space $\H$, the obvious conjugate model is just that associated with the conjugate Hilbert space $\overline{\H}$, with $\gamma_{A}$ taking the rank-one projection $x$ to the corresponding projection $\bar{x}$ on $\overline{\H}$, and with $\eta_{A}$ the 
pure state associated with the unit vector $\frac{1}{\sqrt{n}}\sum_{i} e_i \otimes \bar{e}_i$, $\{e_i\}$ any 
basis for $\H$ (note that this is basis-independent). 

Returning to the general case, note that by averaging over $G(A)$, we can choose $\eta_{A}$ to be invariant, in 
the sense that $\eta_{A}(gx, \gamma_{A}(gy)) = \eta_{A}(gx, gy)$ for all $g \in G(A)$. This gives us an invariant SPIN form on $\E(A)$, defined on outcomes by $B(x,y) := \eta(x,\gamma_{A}(y))$. Applying Theorem~\ref{thm-1}, we have 

\begin{theorem}\label{thm-2} Let $\E(A)$ be irreducible, and suppose $A$ has a conjugate. Then $\E(A)$ carries a 
canonical orthogonalizing inner product.
\end{theorem}

Under some mild auxiliary hypotheses, the existence of a conjugate for every $A \in \C$ (with $\gamma_A$ and $\eta_A$ appropriately belonging to $\C$) can be used to construct a dagger on the category ${\cal E} \supseteq \E(\C)$ discussed above. In fact, however, the mere existence of a reasonable dagger-monoidal structure on ${\cal E}$ is enough to obtain much the same result.


\begin{theorem}\label{thm-3} Suppose ${\cal E}$ supports a dagger-monoidal structure, such that for 
every $g \in G(A)$, $g^{\dagger} = g^{-1}$ (i.e., $g \in G(A)$ is ``unitary"). If $\E(A)$ is irreducible, 
then it carries an orthogonalizing inner product.
\end{theorem}

In order to obtain the self-duality of $\E(A)_{+}$ for an irreducible model $A$, it now suffices to assume either 
of two simple further conditions: 

\begin{theorem} Suppose that either
\begin{itemize}
\item[(a)]  In the context of Theorem~\ref{thm-2}, $A$ has a conjugate such that the state $\eta_{A}$ is an {\em isomorphism state} or 
\item[(b)] In the context of Theorem~\ref{thm-3}, $A$ is {\em sharp}, meaning that every outcome has probability one in a unique state on $\E(A)$. 
\end{itemize} 
Then $\E(A)_+$ is self-dual.
\end{theorem}

The homogeneity of $\E(A)_+$ can now be enforced by any of several conditions discussed 
in \cite{BGW, W11}. Applying the Koecher-Vinberg Theorem, we can conclude that $\E(A)$ carries a unique Jordan 
product making it a formally real Jordan algebra.  

One of these conditions is so simple it's worth pausing to describe it. 
Any bipartite state $\omega$ between $A, B \in \C$ gives rise to a natural positive linear mapping $\hat{\omega} : \E(A) \rightarrow \E(B)^{\ast}$, 
uniquely defined by $\hat{\omega}(x)(y) = \omega(x,y)$. Where $\hat{\omega}$ is an {\em order-isomorphism} --- 
that is, where $\hat{\omega}$ is an order-isomorphism (that is, invertible and having a positive inverse), we 
call $\omega$ an {\em isomorphism state}. A basic observation from \cite{BGW}, translated into the present 
context, is that if every state in the 
interior of $\Omega(B)$ is the marginal of an isomorphism state, then the cone in $\E(B)$ generated by 
$\Omega(B)$ is homogeneous. 

\section{Image-Closure} 

In order to extend these results to possibly reducible systems, I impose one further constraint on $\C$. Call a morphism $(\phi,\psi) : (X,\A,\Omega,G) \rightarrow (Y,\B,\Gamma,H)$ is {\em surjective} iff $\phi(X) = Y$, $\B \subseteq \phi(\A)$, $H = \psi(G)$, and 
$\Gamma = \{ \beta \in \Omega(Y,\B) | \phi^{\ast}(\beta) \in \Omega\}.$
In this case, we call $(Y,\B,\Gamma,H)$ the {\em image} of $(X,\A,\Omega,G)$ under $(\phi,\psi)$. 
Call $\C$ {\em image-closed} iff, for any $A \in \C$ and any surjective morphism $(\phi,\psi) : (X_A,\A_A,\Omega_A,G_A) \rightarrow (Y,\B,\Gamma,H)$, 
(i) the model $B := (Y,\B,\Gamma,H)$ belongs to $\C$, and 
(ii) $(\phi,\psi) \in \C(A,B)$. in $\C$, again belong to $\C$.  

\begin{theorem} Let $\C$ be an image-closed category of bi-symmetric probabilistic models, and let $\cal E$ be the corresponding linearized category 
as discussed in Section 1. If either 
\begin{itemize} 
\item[(a)] every $A \in \C$ has a conjugate $\overline{A} \in \C$, with $\eta_{A}$ an isomorphism state, or 
\item[(b)] $\cal E$ has a dagger-monoidal structure making every $g \in G(A)$ unitary for all $A \in \C$, and every $A \in \C$ is sharp, 
then for every $A \in \C$, $\E(A)_{+}$ is self-dual. \end{itemize}
\end{theorem}

Again, adding any of the sufficient conditions for homogeneity from \cite{BGW,W09} --- or simply assuming 
it outright ---  will yield a category of formally real Jordan algebras. 

Operationally, it is reasonable to suppose that any image $\phi(A)$ of a model $A \in \C$ can be {\em simulated} by 
means of the model $A$. Hence, if we wish to think of $\C$ as closed under operationally reasonable constructions, it is not far-fetched that $\phi(A)$ should belong to $\C$. In fact, the image of a bi-symmetric 
model is $2$-symmetric, so one can simply ``close up" $\C$ without sacrificing this assumption. (To suppose that 
this closure continues to support, e.g., a symmetric-monoidal structure, or conjugate systems, is a sharper 
constraint, of course.) Categories of finite-dimensional quantum models turn out to be image-closed for the simple reason that a quantum model {\em has no} non-trivial images.

\section{Conclusion} 

These results raise any number of interesting questions. For one thing, it is possible that the assumptions are stronger than advertised, singling out a narrower class than formally real Jordan algebras.  It is noteworthy 
that I have not had to assume that $\C$'s monoidal product is locally tomographic. In fact, in a forthcoming 
paper with Howard Barnum \cite{BWta}, we show (using a result of Hanche-Olsen) that if $\C$ is a dagger-monoidal category of finite-dimensional order-unit spaces with homogeneous self-dual cones, then local tomography, plus the existence in $\C$ of a system having the structure of a qubit, implies that every $A \in \C$ is isomorphic to the 
Hermitian part of a finite-dimensional complex $C^{\ast}$ algebra. 


\end{document}